# WS$_2$ Monolayer Integration in a FAPbI$_3$-based Heterostructure

Jorge Luis Miró-Zárate[1], Felipe Cervantes-Sodi[*,2], Milton Carlos Elias-Espinosa[1], Skarleth García-Trujillo[2], Carolina Janani Diliegros-Godines[*,3]

[1]Instituto Tecnológico y de Estudios Superiores de Monterrey, Campus Santa Fe, School of Engineering and Science, Mexico City 01389, Mexico

[2]Physics and Mathematics Department, Universidad Iberoamericana Ciudad de México, Prolongación Paseo de la Reforma 880, Mexico City 01219, Mexico

[3]Instituto de Física, Benemérita Universidad Autónoma de Puebla, Edificio lF1 Ciudad Universitaria, Puebla, Pue, 72750, Mexico

*Corresponding author: felipe.cervantes@ibero.mx and carolina.diliegros@correo.buap.mx

**Abstract**

Incorporating a monolayer of WS$_2$ via interface engineering enhances the overall physical properties of a FAPbI$_3$ perovskite based heterostructure. FAPbI$_3$/WS$_2$/TiO$_2$/ITO and FAPbI$_3$/TiO$_2$/ITO heterostructures were analyzed by UV-Vis spectroscopy, X-ray diffraction, scanning electron microscopy and atomic force microscopy. The configuration with WS$_2$ interlayer presents higher absorption in the visible region with a bandgap of ~1.44 eV. WS$_2$ also enhances the deposition process of FAPbI$_3$, resulting in the formation of pure photoactive α-phase without the non-photoactive δ-phase nor residual plumbates. The incorporation of the monolayer improves the crystalline structure of the FAPbI$_3$, promoting a preferential growth in the [100] direction. The smooth surface of WS$_2$ favors a homogeneous morphology and an increase of the grain size to ~4.5 μm, the largest reported for similar structures. Furthermore, the work function obtained lets us propose an enhance an adequate energy band alignment between FAPbI$_3$ and the n-type layers for the electron flux to the cathode. These findings strongly suggest that the interfacial coupling of FAPbI$_3$/WS$_2$ could be a promising candidate in photovoltaic applications.

The necessity to enhance the photovoltaic properties of perovskite based heterostructures has prompted the incorporation of new layers and functional materials into its the original architecture. A typical perovskite architecture comprises of HTL/perovskite/ETL/TCO/Glass [1,2,3], where TCO, ETL, and HTL are the transparent conductive oxide, electron transport layer, and hole transport layer, respectively. The above-mentioned heterostructures are known to be affected by trap states and interfacial defects, resulting in a loss of electronic and optoelectronic properties due to inadequate interfacial charge transfer[4].

Recently, 2D semiconducting materials, such as transition metal dichalcogenides (TMDs), in particular their monolayers (ML), have been reported to improve the charge separation mechanism, reduce the interfacial recombination and passivate defects, and improve the energy level alignment when incorporated into a heterostructure[5]. H phase $WS_2$ present an electron mobility of ~28 $cm^2$ $V^{-1}$ $s^{-1}$ [6], a direct energy band gap ($E_g$) as ML and inherent n-type semiconductor characteristics[7]. It has been studied in 2D van der Waals heterostructures, and in 2D-3D heterostructures with photoactive materials.[8] Density Functional Theory (DFT) calculations suggests that the incorporation of ML $WS_2$ in a photoactive structure, as ETL, can improve the performance of power conversion efficiency (PCE) [7].

Among photoactive materials, hybrid inorganic-organic perovskite such as $FAPbI_3$ exhibits outstanding optoelectronic properties and a $E_g$ of 1.42 eV [9], optimal for photovoltaic applications. Nevertheless, $FAPbI_3$ properties depend not only upon the perovskite itself, but also, on its interaction with adjacent layers, their quality, and the interfaces [8].

Compact layers of $TiO_2$ are the most used ETLs in perovskite-based heterostructures[10]. However, $TiO_2$ displays a low electron mobility of 0.11-4.15 $cm^2$ $V^{-1}$ $s^{-1}$ [11,12], and inadequate charge separation with the perovskite layer[13]. Promising approaches for optimization the ETL have been explored to enhance its properties, such as interface engineering, through the incorporation of materials like TMDs at the ETL/perovskite interface[14].

In this work, the structural and optoelectronic properties of $FAPbI_3$/$TiO_2$/ITO/Glass and $FAPbI_3$/$WS_2$/$TiO_2$/ITO/Glass heterostructures are presented. A polystyrene (PS) assisted transfer methodology was employed to incorporate the $WS_2$, grown initially on sapphire, into the $TiO_2$ layer for its subsequent use in a perovskite-based heterostructure by depositing the $FAPbI_3$ photoactive layer on top of the $WS_2$. All the experiments were performed under an ambient air atmosphere. It was demonstrated that the incorporation of $WS_2$ as an interlayer promotes the α-phase of $FAPbI_3$, with no trace of the δ-phase of the $FAPbI_3$ nor $PbI_2$. The effect of the $WS_2$ MLs in $FAPbI_3$ surface, results in homogeneous morphology and free-pin holes surface with grain size of ~4.5 μm, enhancing the absorption in the visible range. Work function (Φ) measurements performed by atomic force microscopy (AFM) prove that the $WS_2$ interlayer enables appropriate band energy alignment for electron flux from the $FAPbI_3$ layer to the ITO. This work reports the potential of the $WS_2$ as an interlayer into a perovskite solar cell (PSC) to achieve highly efficient devices.

A modified chemical vapor deposition (CVD) method[15] was used to grow $WS_2$ on C-plane sapphire substrates. 30 ml of 0.02M solution of $Na_2WO_3$ was spin coated on 0.8x0.8 cm² substrates at 3.5krpm for 1 minute and loaded in the center of a 1in quartz tube that was positioned through a tubular furnace. 100mg of S powder was placed inside the tube, 20cm upstream of the substrates. A 15-minute 200sccm Ar purge preceded the growth process in the S and W precursors, which were heated to 220°C and 825 °C with an Ar flux of 100sccm. At these temperatures, 10sccm $H_2$ flux was added for 5 minutes, and then the system was allowed to cool down to room temperature.

A wet transfer method was employed to delaminate the $WS_2$ films on sapphire[16]. An aqueous PS solution was prepared and spin-coated at 3.5krpm, followed by an annealed treatment at 85°C. A water droplet method was employed to form pathways to penetrate the $WS_2$/sapphire interface. The detached film was then slightly lifted off from the sapphire and transferred to the $TiO_2$/ITO/Glass, followed by thermal treatment. PS was removed by dissolving in toluene. The $TiO_2$ film was grown following the methodology reported in reference[17].

$FAPbI_3$ films were obtained from a modified methodology previously reported by the authors[18]. 1 M precursor solution of $PbI_2$ and $NH_4SCN$ were mixed in DMF and spin-coated on the $TiO_2$/ITO/Glass and $WS_2$/$TiO_2$/ITO/Glass substrate. Then, the $PbI_2$-coated substrates were immersed in a solution of $CH_5IN_2$ with isopropyl alcohol for 90s, followed by an annealed treatment at 170°C.

A representative scheme of the heterostructure fabrication is presented in Figure 1(a) and the schematic diagram of the resulting $FAPbI_3$ heterostructure is displayed in Figure 1(b).

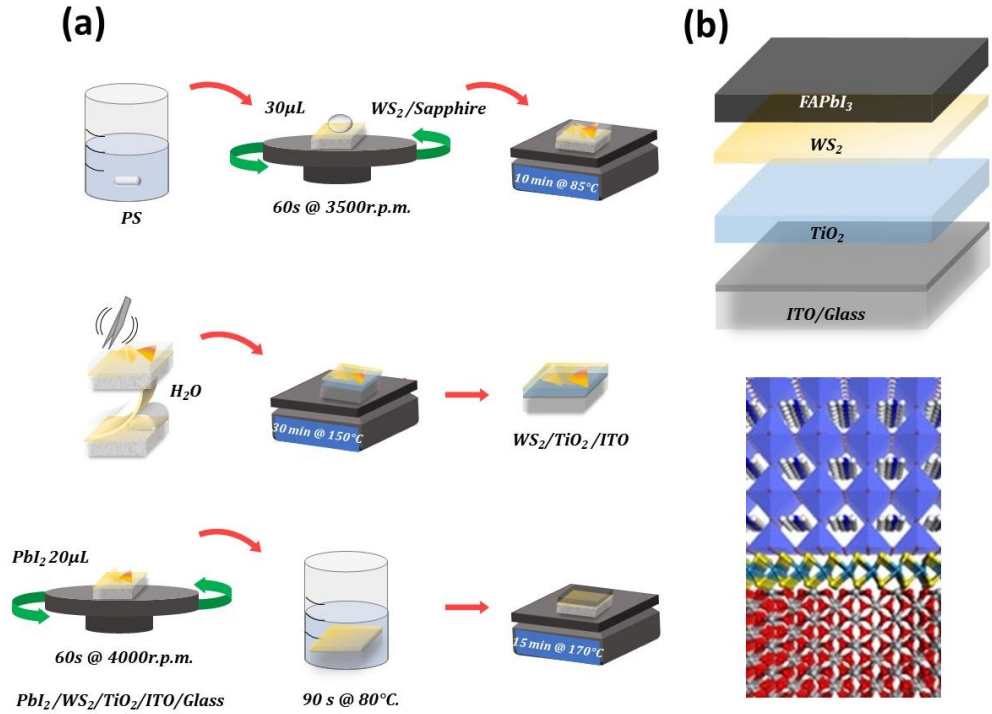

**Figure 1.** (a) Schematic diagram of the wet transfer method of WS$_2$ from a sapphire substrate into the TiO$_2$/ITO/Glass structure, followed by the deposition of FAPbI$_3$ on the heterostructure. (b) Illustration of the FAPbI$_3$/WS$_2$/TiO$_2$/ITO/Glass heterostructure (top) with an atomistic diagram of the FAPbI$_3$/WS$_2$/TiO$_2$ (bottom).

Optical characterization of the heterostructure was performed by ultraviolet-visible-near infrared (UV-Vis-NIR) spectroscopy (Varian Agilent Cary 5000). Structural characterization was conducted by X-ray diffraction (XRD) (D8 Advanced Eco) with CuKα line (1.5418 Å) as X-ray source. Morphology studies were conducted by scanning electron microscopy (SEM) (Hitachi SU 3500) with an accelerating voltage of 5.0 kV.

The work function (Φ) of the individual layers of WS$_2$, TiO$_2$, and FAPbI$_3$ were measured with frequency modulation Kelvin probe force microscopy (KPFM) using sideband configuration following the methodology reported in reference[19]. The probe used was a Multi75-G with a resonance frequency of ~75kHz and a force constant of 3N/m. WS$_2$ individual topography was characterized by AFM NanoSurf model NaioAFM in static mode with a cantilever Stat0.2LAuD.

XRD patterns of the heterostructures are shown in Figure 2 (a). Both heterostructures reveal the characteristic peaks of α-FAPbI$_3$ at 13.9°, 19.8°, 24.3°, 28.1°, 31.5°, 34.5°, 40.2°, 42.8°, 49.7°[20]. Also, ITO peaks are present as expected at 30.4° and 35.3°. Diffraction peaks at 14.2° and 28.8° correspond to the WS$_2$ (JCPDS Card No. 98-003-9096) are only observed for the heterostructure with WS$_2$ interlayer. Remarkably, when the WS$_2$ is incorporated into the heterostructure a preferential growth α-FAPbI$_3$ in the [100] direction is highly promoted.

During the synthesis of FAPbI$_3$ at different conditions, some plumbates could be present due to an incomplete conversion of the precursor PbI$_2$ to the α -phase. The direct growth of FAPbI$_3$ on TiO$_2$ shows a diminished presence of those compared with our previous work[18], where the perovskite was grown directly on glass. Here, the signal of the PbI$_2$ (001) plane at 12.6° disappears when FAPbI$_3$ is grown on WS$_2$. The disappearance of plumbates was also confirmed with SEM images as shown in what follows. Pure α -phase of the FAPbI$_3$ perovskite is grown in both substrates, with and without WS$_2$, which means that both, TiO$_2$ and WS$_2$/TiO$_2$ are appropriate substrates to avoid the formation of the photoinactive δ-phase.

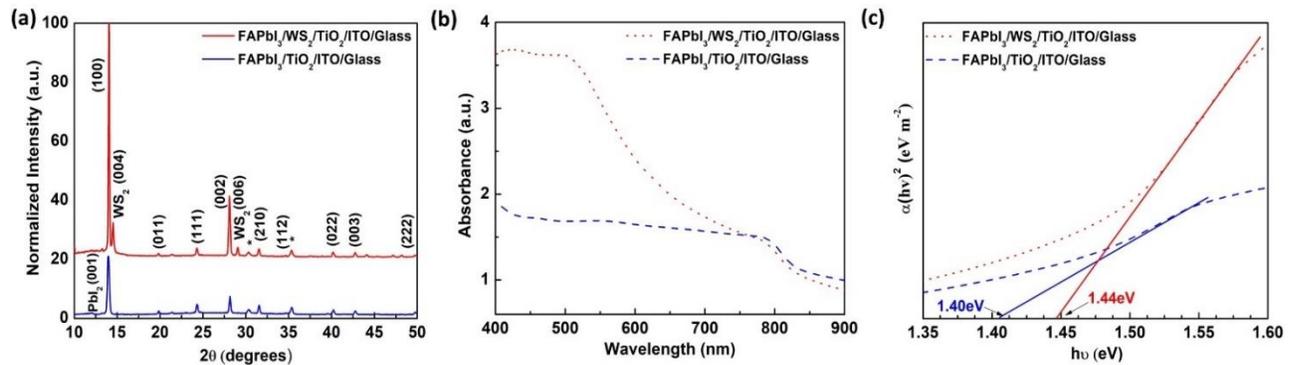

**Figure 2.** (a) XRD patterns of FAPbI$_3$/TiO$_2$/ITO/Glass (bottom) and FAPbI$_3$/WS$_2$/TiO$_2$/ITO/Glass (top) heterostructures, (b) absorbance spectra of heterostructures with and without the WS$_2$ interlayer. (c) Tauc-Plot used to calculate an E$_g$ of 1.44 eV and 1.40 eV with and without WS$_2$, respectively.

The absorption spectra of both heterostructures are shown in Figure 2(b), presenting the characteristic absorption edge of the α-FAPbI$_3$ near to ~800 nm[21]. It is important to point out that the absorption edge of the heterostructure is very similar with and without the WS$_2$ interlayer, indicating that the WS$_2$ does not induce a significative change in the excitation absorption edge. This suggests a type-II band alignment in the FAPbI$_3$/WS$_2$ interface[8]. The presence of the WS$_2$ interlayer slightly increases the absorbance in the visible region, which has been reported when a WS$_2$ layer is used in a CH$_3$NH$_3$PbI$_3$-perovskite based heterostructure[22]. The heterostructure with WS$_2$ shows a flat profile from ~400 to ~500 nm, indicative of fewer structural defects and the presence of single crystalline phases on the FAPbI$_3$ film[23], and promoting the α over the δ phase.

E$_g$ calculations by the Tauc Plot method, as shown in Figure 2(c), provide values of 1.40 and 1.44 *eV* without and with WS$_2$, respectively. The E$_g$ shows a 2.9% increase with the WS$_2$ interlayer. A widening of the E$_g$ can be explained by the atomic thickness of WS$_2$ and a strong dipole-dipole repulsion in the FAPbI$_3$/WS$_2$ interface[24]. E$_g$ of the individual layers were also obtained with the same methodology, with values of 3.04, 1.86, and 1.42 eV for TiO$_2$, WS$_2$, and FAPbI$_3$, respectively, which agrees with the reported in the literature[25].

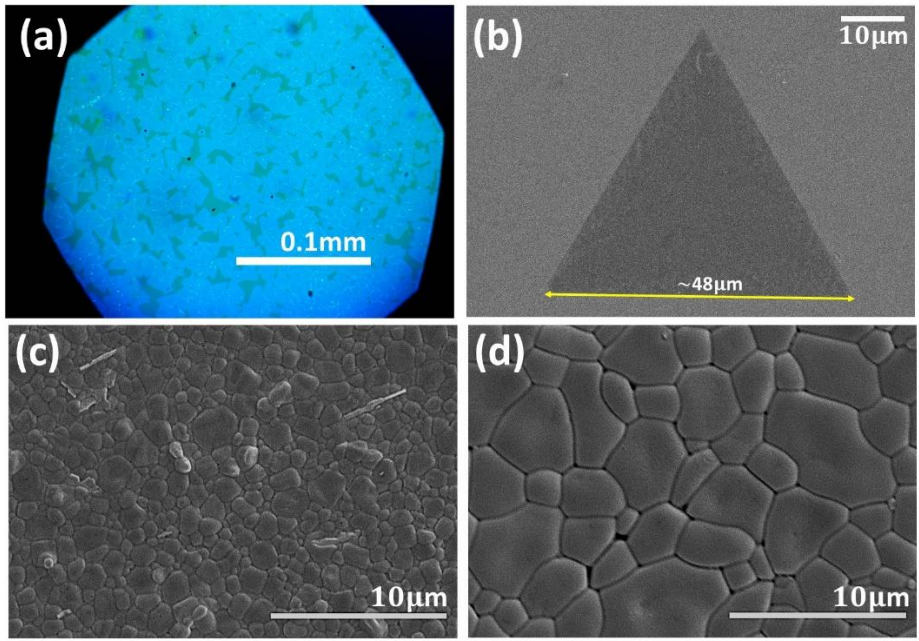

**Figure 3.** (a) Optical image of as grown ML WS$_2$ on sapphire. (b) SEM image of an isolated WS$_2$ ML transferred to TiO$_2$. (c, d) SEM images of FAPbI$_3$ film grown on TiO$_2$/ITO/Glass and WS$_2$/TiO$_2$/ITO/Glass, respectively.

In Figure 3(a), it can be observed the high coverage (> 90%) of WS$_2$ ML, mostly with triangular and hexagonal shapes. The transfer process of WS$_2$ from sapphire to TiO$_2$/ITO/Glass allows an almost perfect WS$_2$ on the ETL, with clean MLs and no PS residues, as can be seen with in a SEM image of an isolated triangular WS$_2$ ML (area ~1000 μm$^2$) transferred to TiO$_2$, shown in Figure 3(b). The high coverage of WS$_2$ on the TiO$_2$ surface results in enhanced FAPbI$_3$ homogeneous morphology and a surface free of pin-holes. Similar to what happens here for FAPbI$_3$, Ma, *et al.* observed that the atomically smooth surface of WS$_2$ promotes high crystallinity of a MAPbI$_3$ film[26].

Figure 3(c) shows that the FAPbI$_3$ perovskite film deposited directly on TiO$_2$ consists of irregular grains of ~1.2μm. The brighter species and irregular shape morphologies over the FAPbI$_3$ surface correspond to multiplumbates generated by an incomplete conversion of PbI$_2$ to FAPbI$_3$; which has been previously reported by the authors[18]. The presence of the multiplumbates was also observed in the XRD patterns, along with the existence of the PbI$_2$ diffraction peak.

The incorporation of WS$_2$ interlayer promotes the growth of larger grains of ~4.5 μm of α-FAPbI$_3$ without any multiplumbates on the film's surface, as illustrated in Figure 3(d). The size of grains in this work are two orders of magnitude larger than other perovskite grains grown on TMDs[27]. Also, a pin-hole free surface can be observed, which can indicate a reduction of trap densities at the material surface [28].

The surface topography height map of WS$_2$ and FAPbI$_3$ and the work function of the WS$_2$, TiO$_2$, FAPbI$_3$ are shown in Figure 4(a-e); all were measured under the same experimental conditions. The surface of the FAPbI$_3$ films is homogeneous, and Φ is almost constant over the surface with Φ ~4.7eV, similar to the value obtained by DFT[29]. The WS$_2$ ML shows a homogeneous surface with roughness ~0.5 nm and thickness > 1 nm. The smooth surface of ML WS$_2$ acts as base for nucleation and growth of large grains for FAPbI$_3$ as previously discussed. Wang, *et al.*, also observed an increase in the grain size of FASnI$_3$ perovskite grown on MoS$_2$, WS$_2$ and WSe$_2$[30]. However, the value of Φ presents some variation across the WS$_2$ surface. Φ increases in the center of the WS$_2$ triangle, showing a defect heterogeneity formation, also reported in the literature[31]. The charge transfer in perovskite-based heterostructures is regulated by the properties of the interfaces of the perovskite with ETL and HTL, the electrochemical reaction during bonding of charged complexes that compensate surface charging, and irreversible material decomposition[32]. However, it is also well known that matching of the Fermi energy ($E_F$) between layers in a device could play a key role in electron recollection. For the heterostructures reported in this work, a smooth match between the TCO and perovskite layer will be reflected in a better device efficiency. Since Φ is the $E_F$ position with respect to the vacuum level, a comparative energy diagram, based on measured Φ and $E_g$ values, is proposed in Figure 4(f). The $E_F$ is pictured as a horizontal dotted line for each film. The $E_F$ for the n-type ETL, ITO and WS$_2$ are shifted to the conduction band[33]. While the FAPbI$_3$ film is located at the center of the gap.

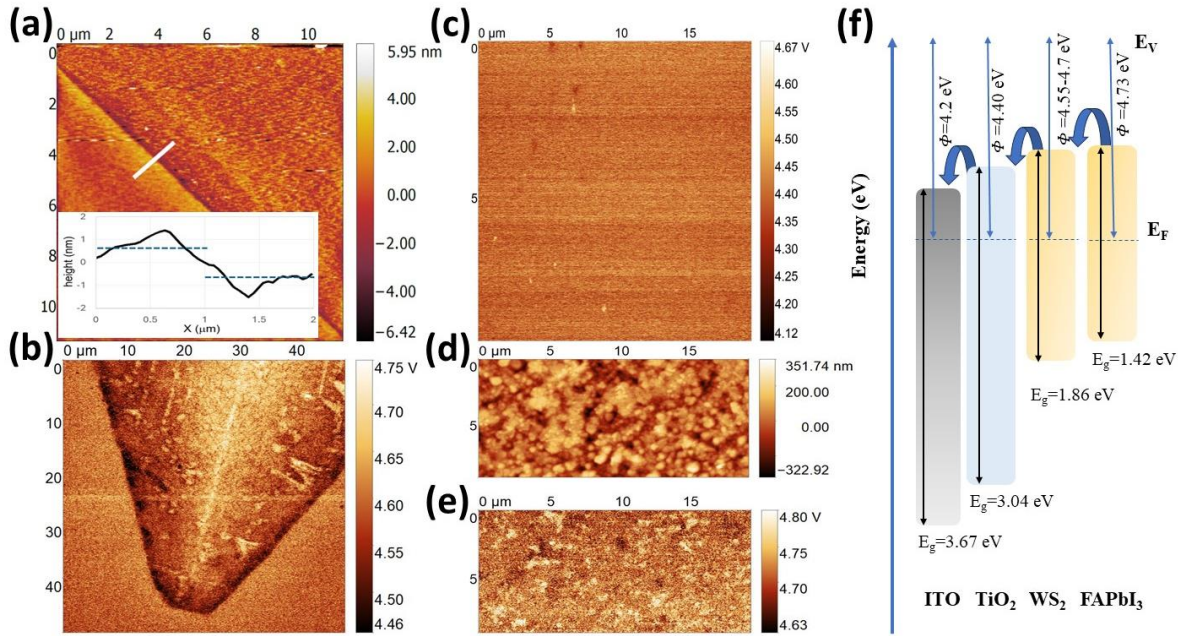

**Figure 4.** (a) Surface topography by AFM of WS$_2$ ML. Inset displays the line scan between the interface WS$_2$/ITO indicated with a white line. Work function values of (b) WS$_2$, (c) TiO$_2$, and (e) FAPbI$_3$. (d) Surface topography of FAPbI$_3$. (f) Comparative energy diagram of the films included in the heterostructure. The values of $E_g$ and Φ were measured.

In Figure 4 (f) ΔΦ between the ITO and ETL is ~0.2 $eV$ and then there is another ΔΦ~0.33eV from the ETL to the perovskite. This gives place to a barrier of energy that the free electron needs to overcome to be recollected. Even when the ETL improves the electron collection the alignment of the $E_F$ observed can provoke charge accumulation and problems of electronic reflux. On the other hand, $\Delta\Phi$ for $WS_2$ shows a smoother transition even with small $\Phi$ variations of $WS_2$ (4.55 to 4.7 eV), than the observed for structures with only the $TiO_2$.

The optimization of $\Phi$ for the layers that compose a perovskite heterostructure has been studied by other authors. Kang Min Kim *et al.*[34] showed that the reduction of $\Phi$ for the transparent layer PEDOT:PSS of the PSC ITO/PEDOT:PSS/$CH_3NH_3PbI_3$/PCBM improves the device performance from 0.07% to 4.02%. Weibo Yan *et al.*[35] studied $\Phi$ of hole-transporting materials (HTMs) in perovskite solar cells, finding that rationally increasing $\Phi$ of HTMs proves beneficial in improving the open circuit voltage of the devices with an ITO/conductive-polymer/$CH_3NH_3PbI_3$/$C_{60}$/BCP/Ag structure.

In this work, the incorporation of $WS_2$ in the heterostructure shows a smoother energy level alignment. This could improve the electron collection, decreasing the recombination process. Therefore, a better performance of the device will be expected, with the improvement of fill factor and short-circuit current density.

In summary, this work presented the experimental incorporation of $WS_2$ in a $FAPbI_3$/$WS_2$/$TiO_2$/ITO/Glass heterostructure. The presence of $WS_2$ ML enhances the optical and morphology properties of the $FAPbI_3$ in the heterostructure. A pure α–phase of the perovskite and the disappearance of plumbates was observed by XRD and confirmed with SEM images for the heterostructures with $WS_2$. SEM images also show grains of ~4.5 μm of α-$FAPbI_3$ on $WS_2$. UV-Vis results suggest a type-II band alignment in the $FAPbI_3$/$WS_2$ interface and a widening of 2.9% of the $E_g$. A smooth energy band alignment was observed by KPFM for the heterostructures with $WS_2$. Our work presents an experimental analysis of the advantages of the $WS_2$ as a complementary ETL in perovskites solar cells. Further analysis should be conducted to confirm the improvement of solar efficiency through the incorporation of $WS_2$.

## Acknowledgements

The JLMZ acknowledges CONAHCYT for support through a doctoral studies scholarship. FCS acknowledges funding and support from DINVP and FISMAT at Universidad Iberoamericana Ciudad de México. SGT acknowledges the "*si quieres puedes*" scholarship from FICSAC, Universidad Iberoamericana Ciudad de México. Part of this work was funded by the CONAHCYT Synergy project number 1564464.

## AUTHOR DECLARATIONS

### Conflict of Interest

The authors have no conflicts to disclose.

### Author Contributions

Conceptualization: JLMZ, FCS, CJDG. Investigation: JLMZ, SGT, FCS, CJDG. Data Curation, Formal Analysis: JLMZ, FCS, CJDG. Funding Acquisition: FCS, CJDG. Resources: FCS, CJDG, MCEE. Supervision: FCS, CJDG. Methodology, Visualization: JLMZ, SGT. Validation: JLMZ, SGT, FCS, CJDG. Writing/Original Draft Preparation: JLMZ. Writing/Review & Editing: JLMZ, FCS, CJDG.

## DATA AVAILABILITY

The data that support the findings of this study are available from the corresponding author upon reasonable request.